\documentclass[journal]{IEEEtran}

\usepackage[ruled]{algorithm2e}

\SetAlFnt{\small}
\SetAlCapFnt{\small}
\SetAlCapNameFnt{\small}
\SetAlCapHSkip{0pt}
\IncMargin{-\parindent}

\usepackage{graphicx}

\usepackage{verbatimbox}
\usepackage{float}
\usepackage{subfig}

\usepackage{fancyvrb}
 
\usepackage{mathtools}

\usepackage{enumerate}
\usepackage{amssymb}
\usepackage{amstext}
\usepackage{amsmath}

\usepackage{subfig}

\begin{document}
 
%

\title{The Internet of Flying-Things:\\Opportunities and Challenges with Airborne Fog Computing and Mobile Cloud in the Clouds
}

%
%
%
%
%

%
\author{
%
%
Seng W. Loke~\IEEEmembership{Member,~IEEE}\\
       \thanks{Seng W. Loke is with the Department of Computer Science and Informationn Technology, La Trobe University, Victoria 3086, Australia, e-mail:
      s.loke@latrobe.edu.au}
       }


\markboth{Internet of Things Journal,~Vol.~X, No.~X, XXX~20XX}%
{Loke}

\maketitle
\begin{abstract} 
This paper focuses on  services and applications provided to mobile users using  airborne computing infrastructure.
We present concepts such as drones-as-a-service and  fly-in,fly-out infrastructure, and note data management and system design issues that arise in these scenarios.
Issues of Big Data arising from such applications, optimising the configuration of airborne and ground infrastructure to provide the best QoS and QoE, situation-awareness, scalability, reliability, scheduling for efficiency, interaction with users and drones using physical annotations are outlined.
\end{abstract}



\IEEEpeerreviewmaketitle

\section{Introduction}

\IEEEPARstart{T}{he}  Google Loon project  captured the world's imagination with a deployment of a network of balloons forming a floating network of
 LTE nodes that can provide rural areas with Internet connectivity, with prototypes already deployed in New Zealand.\footnote{http://www.google.com/loon/}
 Facebook also has a strategy to deploy satellites and drones to provide Internet connectivity in rural areas.\footnote{http://internet.org/}
 Certainly, cloud or cloudlet servers in the air is not too far a stretch of the imagination to consider, but coming with its own data management issues.

Fog networking refers to network architecture that ``uses one or more end-user clients or {\em near-user edge devices} to carry out a substantial amount of storage (rather than stored primarily in cloud data centers), communication (rather than routed over backbone networks), and control, configuration, measurement and management (rather than controlled primarily by network gateways such as those in the LTE core).''\footnote{http://fogresearch.org/about/}  Indeed, fog computing~\cite{fogcomp14} relates to cloudlet computing~\cite{Satyanarayanan:2013:CLE:2465478.2465494,Verbelen:2012:CBC:2307849.2307858} and  mobile cloud computing~\cite{Fernando:2013:MCC:2388122.2388260,fitzekkatz14} and computing with crowds of local near-by devices~\cite{Loke:2015:MCS:2737797.2656214}.

 This paper considers scenarios and systems in fog computing involving {\em near-user edge devices} which are flying or hovering in the air, i.e. literally what we call {\em airborne fog computing}, as well as where mobile device users are provided with services delivered via flying devices such as drones, also known as Unmanned Aerial Vehicles (UAVs).

 The aim of this paper is to highlight a range of scenarios for such airborne fog computing and then to outline a range of challenges.

\section{Scenarios}

\subsection{Service Drones and the Internet-of-Flying-Things}
US Drones are in popular press for combat and surveillance but there has been recent initiatives to adapt successful technology for homeland security.
For example,  Congressman McCaul has helped to ``secure three Unmanned Aerial Vehicles for Texas to provide law enforcement better border surveillance''.\footnote{http://homeland.house.gov/about/chair; http://homeland.house.gov/hearing/\\subcommittee-hearing-using-unmanned-aerial-systems-within-\\homeland-security-game-changer}
Hence, peacetime and civilian users of drones are being proposed but not without its own challenges such as  GPS spoofing on drones that rely heavily on GPS for flight control,\footnote{http://homeland.house.gov/sites/\\homeland.house.gov/files/\\Testimony-Humphreys.pdf} and privacy regulatory issues.

There are also interesting proposals for using drones for security (e.g.,~\cite{safecities15}); for example, drones can be called in on-demand to accompany individuals as they walk home alone at night, perching at particular hotspots or charging points within a park while watching out for the person, or drones may be called in to patrol the house of a disabled or elderly person, who simply cannot afford advanced security cameras and services. Or someone who is leaving home for a holiday for a few days might call in drones to keep an eye on the house. The drones' camera feeds might sent to a security service station for monitoring or image processing technology might be employed to issue warnings when certain persons or intruders are detected.

Drones have found their way into popular press and are gradually finding their way into a range of applications, including filming and surveying,\footnote{http://www.service-drone.com/,\\http://www.auav.com.au}  delivering beer, books and medical supplies such as vaccines and first-aid kits, especially across remote land areas,\footnote{http://matternet.us/,\\http://edition.cnn.com/2013/10/18/tech/\\innovation/zookal-will-deliver-textbooks-using-drones/,http://www.techthefuture.com/technology/\\dronenet-decentralized-delivery-system-of-flying-robots/,http://www.wired.com/tag/uav/}
helping the disabled,\footnote{http://www.dailymail.co.uk/sciencetech/article-2196407/The-flying-quadcopter-disabled-people-control-mind--use-virtual-eyes.html} farming,\footnote{http://www.agweb.com/article/eight\\\_ways\_to\_employ\_drones\\\_on\_the\_farm\_NAA\_Ben\_Potter/} and emergency 
services.\footnote{http://www.heraldsun.com.au/news/victoria/drones-may-be-used-to-help-battle-bushfires-this-summer/story-fni0fit3-1226709860197, http://www.theguardian.com/world/2013/aug/17/florida-keys-drones-mosquito}
 Conservation drones have also been developed that can automatically detect people and animals\footnote{http://conservationdrones.org} and collect footage of wildlife and forests over vast areas.
 
Meanwhile, fascinating demonstrations of robot drones have captured world-wide attention, demonstrating continued developments in and new possibilities for improving the mobility and control of such drones, such as 
 coordinated high precision flying,\footnote{https://droneconference.org/2013/08/vijay-kumar-opening-keynote-at-darc/} 
 creatively constructing a structure,\footnote{http://www.dezeen.com/2011/11/24/flight-assembled-architecture-by-gramazio-kohler-and-raffaello-dandrea/}
 and the joggobot.\footnote{http://exertiongameslab.org/projects/joggobot}  

  
Other work have looked into computer vision control of smart drone navigation,\footnote{http://vision.in.tum.de/data/software/tum\_ardrone} and perhaps it would not be difficult to equip drones with deep learning capabilities.
Indeed, drones are seen to have a potential market, not just for military-related use but for consumer markets, business, and non-commercial operations.
With wide-spread use of mobile devices, the combination of mobile devices and drones (and robots in general) provide an interesting relatively new service delivery platform.

While people might have their own personal drones that they can use out of their pockets, regulatory, practical,  and technical constraints might force the usage of drones into particular service niches. Drones might provide services used by a number of people or people might pay for drones that can deliver a particular type of service, i.e., the idea of drones-as-a-service~\cite{kdrones} rather than requiring every person necessarily deploying their own network of drones.  Moreover, such service drones might be controlled to move along pre-specified paths and routes rather than being able to move chaotically or freely within the airspace of a given area.

Drones may be used to guide people, e.g., for tourists or in emergency scenarios, or they may be used for photo-taking; people in large tourist areas might want photos or videos to be taken from ``impossible angles'' such as off the edge of the cliff looking in, or avid surfers might want their pictures taken while surfing, from different angles, and not just from GoPro perspectives. The idea of drones-as-a-service is that a mobile user (e.g., a tourist or surfer) might place an order for such drones to come to their locations via their mobile devices, and thereafter, be given partial control over the drones via their mobile users (e.g., they can only move the drones within certain zones and have restricted control over drone functions).

 There have also been suggestion for the use of drones to deliver services to the disabled; someone confined to a wheel chair might not mind an avatar in the form of a drone that can be used to ``hang out'' with friends remotely. Also, such a drone can be deployed on the fly and act as ``eyes'' (via streaming video) and ``ears'' (via streaming audio) for the disabled person (e.g., blind or deaf), aided by a crowd of volunteers or friends who convert scenes into voice, and sounds into text and pictures, in the sense of VizWiz.\footnote{http://www.vizwiz.org}

 While the computational, sensing, storage and networking capacities of drones will dramatically improve in the years to come, 
 many of the applications above will involve massive data requirements, especially when video is involved, and especially with increasingly high quality cameras. 
 Such drones will stream videos to other larger drones or to base stations for further storage and processing and might perform a certain amount of processing on-board. 
 
 The constraints of energy and battery will continue to affect what such drones can do, as well as affect the scheduling needs of a set of drones, e.g., where drones periodically replace one another in continuously monitoring work. While relatively cheap hobbyist drones such as the Parrot AR.Drone 2.0\footnote{http://ardrone2.parrot.com} might have around 15-25 minutes of flight time, larger drones  that can stay in the air for up to months and years based on solar power and other energy sources are emerging, but  at huge costs. Drones that combine engines with balloon capabilities (similar to {\em airships} in the past, but on a smaller scale) might emerge, but energy will continue to be a consideration in such drone applications, especially in long term continual monitoring applications. There are challenges of devising suitable schemes for long term monitoring within constraints of costs (which restrict the type and capacity of drones that can be employed) and quality of monitoring (affecting storage, processing, and transmission requirements).

 There are also interesting networking issues for transfer of data among drones and floating cloud servers which need to happen in short bursts when the nodes are within short communication range (to save energy) for short periods and at high transmission rates.  A drone flying among a network of floating cloud servers might need to determine the best collection of cloud servers to offload data to, minimising time and energy constraints -  such drones needs to quickly connect, transfer and go.
 Mini-drones but with networking capabilities but limited storage might be able to transmit their data to larger drones or other airborne servers while in operation, even perhaps recharging from much larger drones and floating base stations, while transferring data.

\subsection{Drones as Data Mules}

The notion of data mules where drones are used to fly near to sensor nodes to collect data thereby saving energy by reducing transmission range, has been presented in ~\cite{Sugihara:2010:SCS:1806895.1806899,Levin:2014:CDA:2592313.2592611}, where the focus is on motion planning of the drones; collected data must satisfy  temporal and spatial (communication range) constraints. Indeed, the idea of drones for collecting data presents opportunities for widely deployed sensor networks. 

Alternative uses could be not just picking up data but drones for picking up dead sensor nodes and replacing such sensor nodes with new ones (or even drones carrying mini-batteries and replacing them in sensor nodes though technically seemingly harder to do so), thereby servicing an entire sensor network. Drones, or balloon-flown nodes, can also function as new intermediate level base stations deployed temporarily and on-demand to minimise transmission sensor node ranges in the case of specific events happening, or provide a basis for sensor networks where each node is mobile (and perhaps can fly) and reconfigure in the presence of new events or in order to redistribute position-dependent responsibilities.

\subsection{Servers in the Air: Fly-In, Fly-Out Infrastructure}

Indeed, as inspired by the Google Loon project, fly-in, fly-out infrastructure, for example, in the case of disaster scenarios, or even for short term large-scale outdoor events (held in remote areas such as a national park or a remote castle, say), is now much closer to realisation. 

For instance,  a balloon suspended cloudlet server is flown temporarily over an area to provide video on-demand services to a group of people for a few days. There are then issues of optimal number and configuration of such a network of servers in the air, that minimises costs, energy usage yet providing adequate quality of experience in terms of providing storage, processing and networking services, for the needs of the event (e.g., the number of users accessing the content from their mobile devices, the type of content the infrastructure is being used for and so on). 

 An advantage of such infrastructure is that they are adaptable and potentially scalable - to service more users than expected or more content than expected, additional cloud servers can be flown to the scene or removed. In the case of bad weather, such infrastructure might literally come down, and need to be suspended and degraded temporarily.

 There has also been work showing that offloading computations can save energy (e.g., ~\cite{5432121}). Indeed, a drone server can be flown to  critical stationary or mobile nodes to provide them with energy-saving services; the critical nodes offload moderate amounts of data and computations to the drone servers over shorter ranges of communication in order to save their energy.

\section{Challenges}
Central to the scenarios above are system and data management issues, apart from drone and airborne device engineering control issues, as well as networking issues. Here, we focus on and summarise  a range of issues and challenges  arising from dealing with data in  the above scenarios.

\begin{itemize}

\item {\em Big Data from airborne fog computing applications}: back in 2009, the US military drones (UAVs) alone (e.g., thousands operating in Iran and Afghanistan) generated 24 years' worth of video if watched continuously.\footnote{http://www.defenseindustrydaily.com/uav-data-volume-solutions-06348/} With the use of drones in civilians services and applications as mentioned above, large volumes of streaming data will be expected. There is a need to deal with  such data, including labelling, processing and analysis. There are also challenges in fusing tens to hundreds of streams of video  data for end-users, using airborne and ground infrastructure in an efficient manner.

\item {\em optimizing for multiple factors}: a key, potentially complex, issue in the above scenarios is optimizing an airborne fog computing architectural solution for energy consumption, monetary costs, efficiency (in data transmission and for required data analysis computations) while still meeting application requirements for Quality-of-Service and Quality-of-Experience.  For example, given a fixed budget, what is the configuration of drone services, airborne fog infrastructure and ground servers that would provide the best means of handling video data for a tourist video capturing application at the Mornington Peninsular in Melbourne, Australia? A company setting out to create such drone-based video capturing services might use a combination of existing cloud providers as well as its own airborne fog computing servers and drones. 

A data mule application using drones might have to trade-off drone storage capacities and time constraints for data collection. A related question is what interleaving of data processing (e.g., creating summaries) and data transmission would fit the application best.

\item {\em situation-awareness}: there is a need for context information to inform the airborne fog computing infrastructure in order to support  adaptations  to the current situation. This ranges form weather changes to changes in the position of mobile users being serviced, e.g., in the case of service drones. Drones can move to positions where servicing clients would be expected to require minimal energy. Context data must be collected from a variety of sensors, and fed to appropriate nodes in the airborne fog infrastructure to inform run-time adaptations.

\item {\em scalability, incremental extensibility, and compositionality of services and servers}: 
service composition in order to process data arising from such applications would be required, including services for data storage, caching,  aggregation, processing, analytics and reasoning, and composed on-demand according to current  needs. The idea is that different nodes might be involved in different stages of processing. For example, consider the scenario of a set of drones roaming an area to video capture the area from different perspectives. The data might then be streamed to nearby servers in the air and eventually relayed to ground cloudlet servers (or sent directly to ground servers where reachable) and eventually perhaps archived on a remote Cloud server. Video streams could also come from ground robots and mobile devices carried by users in the field. 

Then, the video aggregation, analysis, synthesis, and processing might happen on the intermediate servers along the way. A suitable configuration of nodes and networking capability in an integrated architecture that optimises transmission costs, time and energy consumption yet fulfilling the requirements of the application is required, and is expected to vary with the area and application-specific requirements. 

Such services should be compositional and the architecture compositional and adaptable, so that, over time, if additional resources (e.g., more drones or more powerful drones) are added to the system, no substantial changes would be required.
If applications are to be built in an incremental fashion (e.g., a drone at a time), and scaled up and out as required, the airborne fog computing infrastructure would then need to be flexible and compositional.

\item {\em reliability}: a key issue is reliability of the system especially when critical data needs to be handled. Redundancy is a common solution; to meet reliability requirements in airborne nodes where many things could go wrong, it might be that a great deal of replication of data is required, increasing costs.  Where and how such data replication happens and at what rate would be a design consideration for an airborne fog computing infrastructure supporting an application. Partitioning of data into grades of required reliability might be useful.

\item {\em scheduling}: online algorithms are required as not all user scenarios can be anticipated. However, analytics on historical service transactions might help inform the scheduling of drones. For example, in the case of the service drones for tourist applications, if it can be found that requests tends to come from certain tourist spots, or in the case of delivery where after some time, frequent routes are mapped, then the drones can be repositioned and scheduled accordingly. In the case of data mules, eventful areas might be visited more often.

\item {\em interaction with and data manipulation services for users}: the user might want not just to capture video streams but also view them on their mobile devices after capture and perhaps even edit their videos on their mobile devices. So, drones that capture such video might stream data to servers but then a mechanism for users  to recover, edit and use such data might also be required. Simple solutions exist where users might just access via the Web their data repositories once they return home but a question is whether more flexible services can be provided, e.g., the user can access their data on-demand even while in the remote area, and even control exactly when and where video is captured by the drones serving them.

\item {\em drones using physical annotation data}: there has been work on augmented reality for users where visual data are superimposed on objects seen through a camera by the user, and there has been work on users leaving geo-tagged notes for other users where the users only see the notes when they are at the right locations~\cite{Hansen:2006:UAS:1149941.1149967,doi:10.1108/17427371111189647}. Such geo-tagged notes can also be used by drones in applications. For example, if one labelled a location as ``the place where I proposed to my wife'', and tell the drone to fly to that place, the system can resolve the  user-friendly phrase into GPS coordinates and direct the drone to move to that location. A layer of semantic labels can therefore overlay areas which can be used by drones that have access to such data about places. 

\item {\em highways-in-the-air}: it is difficult to conceive of drones flying haphazardly in urban sky; but perhaps drones will be programmed to move within  specific approved pathways or ``virtual tunnels'' in the sky and/or underground. A network of such drones, accompanied by a network of base stations (for charging and delivery),
of different shapes and sizes, suited to their tasks might be employed, but key developments will be needed to implement such virtual tunnelling and constrained movements.

\item {\em situational and context-aware drone control}: contextual control and management of drones could be useful. For example,
at different times, a drone might be able to follow a person, but at another time, they are not allowed to do so. At certain times, a drone might be allowed to enter the airspace of a house, but at other times, they are not. At certain times, control of a drone might be passed to a customer for use, but once rental has expired, control must then be passed back to the owner. 

The ability to change control,  track and localise such drones to ensure that they move within constrained virtual ``boxes'' and pathways depending on context (the time and situation) will need further thought and development. 

\item {\em drone-user interaction}: in the Internet of Things, while automation is enabled, there are still issues of where the human fits in, the human in the loop issue~\cite{6774858}; a similar issue is observed where autonomous drones should share control with humans, so that humans are not burdened by the need to fly and control drones, especially, a fleet of such drones, but yet drones must be able to be programmable by users at a high level of abstraction. Programming robots to perform particular tasks is non-trivial, especially, if end-users need to do so, e.g., to instruct robots or drones to do certain tasks, instead of just predefined tasks. While exciting experiments in brain-controlled drones\footnote{E.g., http://www.bbc.com/news/technology-31584547 and https://www.yahoo.com/tech/watch-this-man-control-a-flying-drone-with-his-112071654249.html} have been newsworthy, appropriate abstractions to interact with drones will be useful, e.g., to instruct a drone to follow someone up to a certain point or to patrol only at certain times and places and only from certain perspectives (for privacy reasons, say).
\end{itemize} 

 \section{Conclusion}
 While drones have been used in the military, there are interesting uses of drones that could be further explored, even as the technology and the ability to control such drones matures.
 Such airborne fog computing infrastructure might provide data to users or collect data for users, or both.
 And there are social-technical and legal issues to be resolved still. 
 
Here, we have outlined a number of scenarios and challenges in providing mobile services via drones 
within   a vision of airborne fog computing.  However, they are, 
by no means exhaustive, and indeed new issues will be uncovered as applications are tested and built in time to come. 


%
\bibliographystyle{abbrv}
\bibliography{kdrone}  

\begin{thebibliography}{10}

\bibitem{doi:10.1108/17427371111189647}
A.~A. Alzahrani, S.~W. Loke, and H.~Lu.
\newblock A survey on internet‐enabled physical annotation systems.
\newblock {\em International Journal of Pervasive Computing and
  Communications}, 7(4):293--315, 2011.

\bibitem{fogcomp14}
F.~Bonomi, R.~Milito, P.~Natarajan, and J.~Zhu.
\newblock Fog computing: A platform for internet of things and analytics.
\newblock In N.~Bessis and C.~Dobre, editors, {\em Big Data and Internet of
  Things: A Roadmap for Smart Environments}, volume 546 of {\em Studies in
  Computational Intelligence}, pages 169--186. Springer International
  Publishing, 2014.

\bibitem{Fernando:2013:MCC:2388122.2388260}
N.~Fernando, S.~W. Loke, and W.~Rahayu.
\newblock Mobile cloud computing: A survey.
\newblock {\em Future Gener. Comput. Syst.}, 29(1):84--106, Jan. 2013.

\bibitem{fitzekkatz14}
F.~H.~P. Fitzek and M.~D. Katz.
\newblock {\em Mobile Clouds: Exploiting Distributed Resources in Wireless
  Networks}.
\newblock Wiley-Blackwell, 2014.

\bibitem{Hansen:2006:UAS:1149941.1149967}
F.~A. Hansen.
\newblock Ubiquitous annotation systems: Technologies and challenges.
\newblock In {\em Proceedings of the Seventeenth Conference on Hypertext and
  Hypermedia}, HYPERTEXT '06, pages 121--132, New York, NY, USA, 2006. ACM.

\bibitem{5432121}
K.~Kumar and Y.-H. Lu.
\newblock Cloud computing for mobile users.
\newblock {\em Computer}, PP(99):1--1, 2011.

\bibitem{Levin:2014:CDA:2592313.2592611}
L.~Levin, A.~Efrat, and M.~Segal.
\newblock Collecting data in ad-hoc networks with reduced uncertainty.
\newblock {\em Ad Hoc Networks}, 17:71--81, June 2014.

\bibitem{safecities15}
S.~Loke and M.~Alwateer.
\newblock Pervasive on-demand safety services in the future: Mobile apps,
  wearable devices, robots, and drones.
\newblock In {\em Proceedings of the Safe Cities Conference (poster paper)},
  July 2015.

\bibitem{kdrones}
S.~W. Loke.
\newblock Smart environments as places serviced by k-drone systems (accepted,
  to appear).
\newblock {\em Journal of Ambient Intelligence and Smart Environments}, 2015.

\bibitem{Loke:2015:MCS:2737797.2656214}
S.~W. Loke, K.~Napier, A.~Alali, N.~Fernando, and W.~Rahayu.
\newblock Mobile computations with surrounding devices: Proximity sensing and
  multilayered work stealing.
\newblock {\em ACM Trans. Embed. Comput. Syst.}, 14(2):22:1--22:25, Feb. 2015.

\bibitem{Satyanarayanan:2013:CLE:2465478.2465494}
M.~Satyanarayanan.
\newblock Cloudlets: At the leading edge of cloud-mobile convergence.
\newblock In {\em Proceedings of the 9th International ACM Sigsoft Conference
  on Quality of Software Architectures}, QoSA '13, pages 1--2, New York, NY,
  USA, 2013. ACM.

\bibitem{6774858}
J.~Stankovic.
\newblock Research directions for the internet of things.
\newblock {\em Internet of Things Journal, IEEE}, 1(1):3--9, Feb 2014.

\bibitem{Sugihara:2010:SCS:1806895.1806899}
R.~Sugihara and R.~K. Gupta.
\newblock Speed control and scheduling of data mules in sensor networks.
\newblock {\em ACM Trans. Sen. Netw.}, 7(1):4:1--4:29, Aug. 2010.

\bibitem{Verbelen:2012:CBC:2307849.2307858}
T.~Verbelen, P.~Simoens, F.~De~Turck, and B.~Dhoedt.
\newblock Cloudlets: Bringing the cloud to the mobile user.
\newblock In {\em Proceedings of the Third ACM Workshop on Mobile Cloud
  Computing and Services}, MCS '12, pages 29--36, New York, NY, USA, 2012. ACM.

\end{thebibliography}
%
%
\end{document}